\documentclass[twocolumn,superscriptaddress,amsmath,amssymb,aps,prx]{revtex4-2}
\usepackage{amsmath,amssymb}
\usepackage{graphicx}
\usepackage{dcolumn}
\usepackage{bm}
\usepackage{multirow}
\usepackage{textcomp}
\usepackage{url}
\usepackage{enumitem}
\usepackage{float}
\usepackage[dvipsnames]{xcolor}
\usepackage[colorlinks=true, allcolors=cyan]{hyperref}
\usepackage{orcidlink}
\newcommand{\bra}[1]{\langle{#1}|}
\newcommand{\ket}[1]{|{#1}\rangle}
\newcommand{\braket}[2]{\langle{#1}|{#2}\rangle}

\begin{document}

\title{\textit{Ab Initio} Exact Calculation of Strongly-Correlated Nucleonic Matter}

\author{R. Z. Hu\,\orcidlink{0009-0002-8797-6622}}
\affiliation{School of Physics, and State Key Laboratory of Nuclear Physics and Technology, Peking University, Beijing 100871, China}
\author{S. L. Jin\,\orcidlink{0000-0001-5958-8070}}
\affiliation{School of Physics, and State Key Laboratory of Nuclear Physics and Technology, Peking University, Beijing 100871, China}
\author{X. Zhen\,\orcidlink{0009-0000-1806-4123}}
\affiliation{School of Physics, and State Key Laboratory of Nuclear Physics and Technology, Peking University, Beijing 100871, China}
\author{H. Y. Shang\,\orcidlink{0009-0007-1253-4519}}
\affiliation{School of Physics, and State Key Laboratory of Nuclear Physics and Technology, Peking University, Beijing 100871, China}
\author{J. C. Pei\,\orcidlink{0000-0002-9286-1304}}\email[]{peij@pku.edu.cn}
\affiliation{School of Physics, and State Key Laboratory of Nuclear Physics and Technology, Peking University, Beijing 100871, China}
\affiliation{Southern Center for Nuclear-Science Theory (SCNT), Institute of Modern Physics, Chinese Academy of Sciences, Huizhou 516000, China}
\author{F. R. Xu\,\orcidlink{0000-0001-6699-0965}}\email[]{frxu@pku.edu.cn}
\affiliation{School of Physics, and State Key Laboratory of Nuclear Physics and Technology, Peking University, Beijing 100871, China}
\affiliation{Southern Center for Nuclear-Science Theory (SCNT), Institute of Modern Physics, Chinese Academy of Sciences, Huizhou 516000, China}
\author{F.~Marino\,\orcidlink{0000-0001-7743-1982}}
\affiliation{Institut f\"{u}r Kernphysik and PRISMA+ Cluster of Excellence, Johannes Gutenberg-Universit\"{a}t Mainz, 55128 Mainz, Germany}

\date{\today}

\begin{abstract}
   Dense nucleonic matter is of vital importance for understanding compact stars and inferring the transition into deconfined quark phase. We present \textit{ab initio} exact calculations of infinite nucleonic matter with the state-of-the-art full configuration-interaction quantum Monte Carlo (FCIQMC) method, enabling us to rigorously benchmark many-body methods and assess the degree to which the nucleonic matter is correlated. Our method has been numerically validated against exact diagonalization within a small model space. Calculations of nucleonic matter using chiral nuclear forces reveal that symmetric nuclear matter is strikingly strongly correlated, raising questions on previous \textit{ab initio} calculations of nuclear matter with many-body expansion truncations and offering insights into simultaneous descriptions of finite nuclei and infinite nucleonic matter from first principles.
\end{abstract}

\maketitle

\textit{Introduction.\textemdash}
{\it Ab initio} nuclear theory has made significant progress
over the past two decades~\cite{10.3389/fphy.2020.00098,10.3389/fphy.2020.00379}, however, exact many-body calculation of infinite nucleonic matter is still a missing piece.
In the \textit{ab initio} nuclear paradigm, starting from realistic nuclear forces, the full-configuration no-core shell model (NCSM) can be seen as exact diagonalization calculations of finite nuclei with active degrees of freedom of all nucleons \cite{BARRETT2013131}. But NCSM is limited to light nuclei up to $^{16}$O owing to formidable computing costs. Moreover, it is essential to examine the applicability of \textit{ab initio} calculations to heavy nuclei considering the in-medium effect~\cite{PhysRevLett.120.152503,PhysRevLett.125.182501,Gysbers2019,BINDER2014119,PhysRevC.105.014302,Hu2024,10.3389/fphy.2019.00213}, although realistic nuclear forces fitted to free-space scattering experiments are verified in few-body systems \cite{PhysRevC.68.041001,PhysRevC.72.034002}. To this end, \textit{ab initio} calculations have to reproduce the empirical properties of infinite nuclear matter at the saturation density~\cite{PhysRevC.96.014303,PhysRevLett.122.042501,PhysRevC.102.054301}.
However, most \textit{ab initio} nuclear methods fail to describe finite nuclei and nuclear matter simultaneously \cite{MACHLEIDT2024104117,HEBELER20211}. From this perspective, \textit{ab initio} exact calculations of infinite nuclear matter are imperative.

The accurate prediction of equation of state (EoS) of cold dense nucleonic matter that is not accessible by terrestrial experiments is vital for  understanding compact neutron stars in the multi-messenger astrophysical era~\cite{annurev-astro-081915-023322,PhysRevLett.126.172503,PhysRevLett.127.192701,PhysRevLett.130.112701,Annala2023}. It also provides a stepping stone for inferring the phase transition scenario from nuclear matter to quark matter, informing us of the possible existence of hybrid stars and non-perturbative properties of strong interactions~\cite{PhysRevLett.124.171103,PhysRevC.108.025803,PhysRevLett.130.091404,PhysRevLett.129.181101,PhysRevD.106.103027}. The theoretical challenge is related to the knowledge that nuclear matter is a strongly correlated system, in particular the symmetric nuclear matter (SNM), while the pure neutron matter (PNM) is relatively weakly correlated~\cite{annurev2015,PhysRevC.101.045801,PhysRevC.105.055808,PhysRevLett.125.202702}. This is indicated by previous studies that different \textit{ab initio} methods have obvious discrepancies in SNM using hard nuclear forces~\cite{PhysRevC.88.044302,PhysRevC.93.065801,PhysRevC.89.014319,PhysRevC.88.054312,PhysRevC.110.054322,ZHEN2025139350,udiani}. Therefore, it will be interesting to see how the truncations applied in \textit{ab initio} calculations of finite nuclei play in infinite nuclear matter.

It is impossible to perform exact diagonalization calculations of infinite nuclear matter by  brute force~\cite{Comput2017}. There are \textit{ab initio} methods with many-body expansion truncations applied to nuclear matter, including the many-body perturbation theory (MBPT) up to the 5th order~\cite{10.3389/fphy.2020.00164,PhysRevLett.122.042501,PhysRevLett.130.072701,drischler2026}, coupled-cluster (CC) truncated at the doubles excitation level with perturbative triple corrections CCD(T)~\cite{PhysRevC.89.014319,PhysRevC.88.054312},
self-consistent Green's function (SCGF) theory~\cite{10.3389/fphy.2020.00387} based on the algebraic diagrammatic construction (ADC) approximation scheme~\cite{PhysRevC.97.054308,PhysRevC.105.044330,PhysRevC.110.054322,marino2026}, and the in-medium similarity renormalization group (IMSRG) truncated to the normal-ordered two-body level~\cite{udiani,ZHEN2025139350}.
While these basis-expansion techniques are highly effective with soft potentials, Quantum Monte Carlo (QMC) approaches are generally advantageous for handling harder interactions. In coordinate space, the auxiliary field diffusion Monte Carlo (AFDMC)~\cite{PhysRevLett.113.182503,PhysRevLett.116.062501,PhysRevC.101.045801} and recent variational calculations using neural-network quantum states~\cite{doi:10.1126/science.aag2302,PhysRevResearch.2.033429,Hermann2020,PhysRevLett.127.022502,KEEBLE2020135743,PhysRevLett.122.226401} have been developed. In configuration space, methods such as the configuration-interaction Monte Carlo (CIMC)~\cite{PhysRevLett.112.221103,PhysRevC.107.044303} have also been available. Although these QMC methods can in principle include high-order correlations, they often suffer from the fermion sign problem or rely on the quality of variational ansatz. To address the aforementioned difficulties, we hereby introduce the state-of-the-art full configuration-interaction quantum Monte Carlo (FCIQMC) method for \textit{ab initio} exact calculations of infinite nuclear matter.

FCIQMC was originally developed in quantum chemistry and quickly has been widely used in molecular and condensed matter physics~\cite{10.1063/1.3193710,PhysRevB.85.081103,Booth2013,10.1063/1.3302277}. FCIQMC has emerged as one of the most accurate many-body methods, particularly for strongly correlated systems \cite{PhysRevB.91.045139,PhysRevX.10.011041}. Different from post-Hartree-Fock methods, FCIQMC is a projector QMC method that stochastically samples the ground-state wavefunction in the full Hilbert space. This approach also offers distinct advantages over other QMC techniques~\cite{doi:10.1021/acs.jctc.8b01217,10.1063/5.0005754}.
Being formulated in a discrete configuration space, unlike coordinate-space methods such as AFDMC, FCIQMC is not restricted to local potentials and can naturally handle nonlocal forces from chiral effective field theory ($\chi$EFT)~\cite{MACHLEIDT2024104117}. Furthermore, FCIQMC employs a delicate walker annihilation algorithm to overcome the fermion sign problem in QMC methods~\cite{10.1063/1.3193710,10.1063/1.3407895}. This contrasts with methods like AFDMC or CIMC, which require the fixed-node approximation or guiding wavefunctions, introducing a bias that is difficult to assess or eliminate~\cite{PhysRevB.55.7464,PhysRevB.102.161104,doi:10.1021/acs.jctc.8b01217,10.1063/5.0005754}. Therefore, as a method that in-principle is exact and possessing well-controlled systematic errors, FCIQMC holds the promise of providing benchmarking and advancing our understandings of dense nuclear matter and nuclear forces.

\textit{Methodology.\textemdash} FCIQMC solves the imaginary-time Schr\"odinger equation~\cite{doi:10.1142/1170},
\begin{equation}
\label{equ:schrodinger}
-\dfrac{\mathrm{d}}{\mathrm{d}\tau} \ket{\Psi(\tau)} = (\hat{H}-E)\ket{\Psi(\tau)},
\end{equation}
where $\hat{H}$ is the many-body Hamiltonian. In the long-time limit ($\tau\to\infty$), the wavefunction $\ket{\Psi(\tau)}$ stochastically projects onto the exact ground state of $\hat{H}$.

The core idea of FCIQMC is to represent the wavefunction as a dynamic population of discrete, signed walkers distributed across the full configuration interaction (FCI) basis of all Slater determinants~\cite{10.1063/1.3193710}, $\ket{\Psi} = \sum_{i} c_i \ket{D_i}$. The coefficient $c_i$ for each determinant $\ket{D_i}$ is sampled by the number of walkers $N_i$ residing on it. The evolution of these walker populations is governed by the master equation:
\begin{equation}
-\dfrac{\mathrm{d}N_i}{\mathrm{d}\tau} = (H_{ii}-S)N_i + \sum_{i\neq j} H_{ij} N_j,
\end{equation}
where $H_{ij}=\braket{D_i}{\hat{H}|D_j}$ are the Hamiltonian matrix elements, and $S$ is an energy offset known as shift which is introduced to prevent the true ground state from decaying to zero. This master equation is simulated via a stochastic algorithm applied at each time step $\Delta \tau$, which comprises three steps~\cite{10.1063/1.4766327}:
\begin{enumerate}
\item \textit{Spawning:} Walkers on a determinant $\ket{D_j}$ can ``spawn" new walkers onto connected determinants $\ket{D_i}$ (where $H_{ij}\neq 0$). The probability of creating a new walker is proportional to the magnitude of the off-diagonal matrix element $|H_{ij}|$.
\item \textit{Death/Cloning:} The diagonal term, $H_{ii}-S$, governs a local process where walkers on $\ket{D_i}$ are removed (death) or duplicated (cloning), depending on whether the local energy $H_{ii}$ is greater or less than the shift $S$.
\item \textit{Annihilation:} Walkers with opposite signs that arrive at the same determinant are removed. This annihilation step is essential for mitigating the fermion sign problem~\cite{PhysRevB.89.245124,doi:10.1021/acs.jctc.1c00078}.
\end{enumerate}

\begin{figure}[t]
\centering
\resizebox{0.48\textwidth}{!}{
\includegraphics{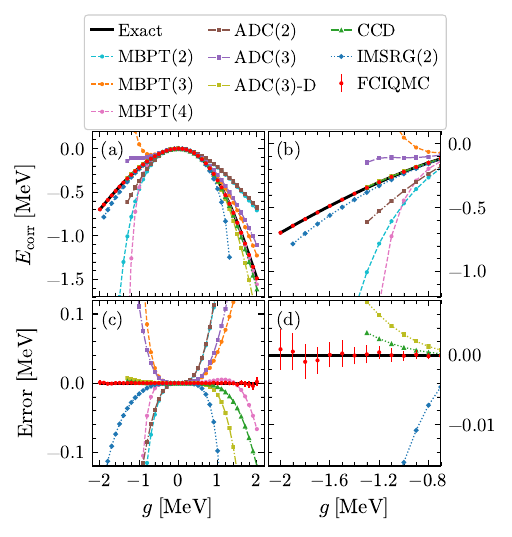}}
\caption{Benchmark results for the Richardson model with $\delta=1.0$ MeV.  (a) and (b) show the correlation energies as functions of the pairing strength $g$. (c) and (d) display the deviations of various many-body methods with respect to the exact solution.}
\label{fig:fig_pairing}
\end{figure}

\begin{figure*}[t]
    \centering
    \includegraphics[width=1\textwidth]{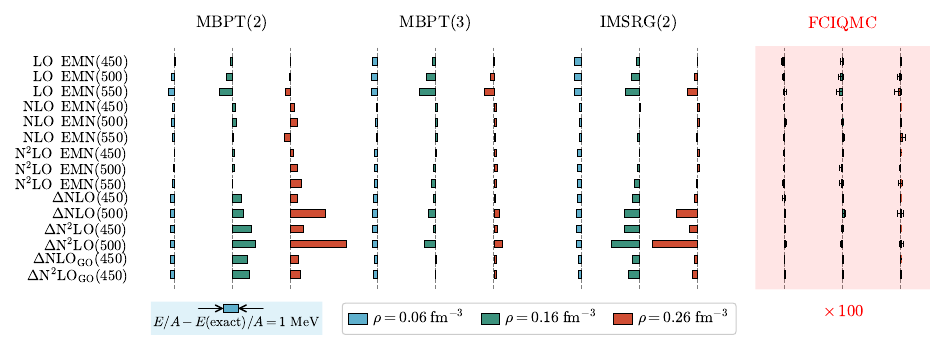}
    \caption{Comparison of many-body methods for SNM in a small model space (4 nucleons, 28 single-particle basis states), benchmarked against exact results at three densities. The bars represent the deviations in the ground-state energy from exact solution for MBPT(2), MBPT(3), IMSRG(2), and FCIQMC. Note that the FCIQMC errors have been magnified by a factor of 100 for visualization, with their statistical uncertainty also shown. Calculations were performed using both $\Delta$-less [$\mathrm{N^\nu LO\; EMN(\Lambda)}$]~\cite{PhysRevC.96.024004} and $\Delta$-full [$\mathrm{\Delta N^\nu LO(\Lambda)}$]~\cite{PhysRevC.97.024332,PhysRevC.102.054301} interactions, in which $\nu$ denotes the chiral order
    and $\Lambda$ is the cutoff (in MeV).}
    \label{fig:comparision}
\end{figure*}

The shift $S$ is dynamically adjusted to maintain a stable total walker number, $N_\mathrm{w}=\sum_i |N_i|$. Once the simulation reaches equilibrium, i.e., $S$ fluctuates around a stable value, the ground-state energy is calculated using the projected estimator,
\begin{equation}
E(\tau)=\dfrac{\bra{D_0}\hat{H}\ket{\Psi(\tau)}}{\braket{D_0}{\Psi(\tau)}}=\sum_{i}H_{0i}\dfrac{N_i(\tau)}{N_0(\tau)},
\end{equation}
where $\ket{D_0}$ is a reference determinant. The computational cost is linearly related to the total walker number.
The ground-state energy and its statistical uncertainty are determined through reblocking analysis~\cite{10.1063/1.457480,pyblock}.

To enhance the efficiency and stability for large systems, we employ the initiator approximation~\cite{10.1063/1.3302277}. This widely-used adaptation designates determinants with a walker population above a certain threshold as ``initiators". Spawning from non-initiator determinants is then restricted to only those determinants that are already occupied. This curtails the proliferation of noise from sparsely populated regions of the Hilbert space. While the initiator approximation introduces a systematic bias, such a bias is progressively reduced by increasing the total number of walkers, allowing for the extrapolation to exact solutions, the unbiased FCI limit $N_\mathrm{w}\to\infty$~\cite{PhysRevB.103.155135,PhysRevB.105.235144}.
The adaptive-shift method~\cite{10.1063/1.5134006,10.1063/5.0005754} has been used to accelerate the $N_\mathrm{w}$ convergence within large model spaces. The full details of the FCIQMC algorithms are shown in the Supplemental Material~\cite{SM}.

\textit{Results.\textemdash}To validate our implementation and demonstrate the powerful capability of FCIQMC, we first benchmark it against several other advanced many-body methods using the exactly solvable Richardson pairing model~\cite{RICHARDSON1963277,RevModPhys.76.643}. This model serves as an ideal testing ground for assessing the accuracy of many-body methods across a wide range of interaction strengths, from weak to strong couplings~\cite{PhysRevE.107.025310,PhysRevC.109.024311,Hjorth-Jensen::LectNotesPhys936,PhysRevLett.134.182502}. The Hamiltonian is given by
\begin{equation}
\hat{H}=\delta \sum_{p=1}^{p_\mathrm{max}} \sum_{\sigma=\uparrow,\downarrow} (p-1)a_{p\sigma}^\dagger a_{p\sigma} -\dfrac{g}{2} \sum_{p,q=1}^{p_\mathrm{max}} a_{p\uparrow}^\dagger a_{p\downarrow}^\dagger a_{q\uparrow} a_{q\downarrow},
\end{equation}
where $\delta$ is the single-particle level spacing (set to 1.0 MeV) and $g$ is the pairing strength. Our benchmark is performed for a system of $A=4$ nucleons in $p_\mathrm{max}=4$ levels (a half-filled case), using a total walker population of $N_\mathrm{w}\approx 10^4$ for the FCIQMC calculations.

Figure~\ref{fig:fig_pairing} presents the correlation energies using different many-body methods and their deviations from the exact solution by direct diagonalzing. As shown, perturbative approaches like MBPT at order $n$ ($n$=2, 3, 4) depart from exact solutions rapidly as the interaction strength $|g|$ increases. While non-perturbative methods such as CCD, ADC-SCGF at third-order in the ADC hierarchy plus CCD amplitude corrections [ADC(3)-D]~\cite{Barbieri2017,marino2026}, and IMSRG(2) offer significant improvements in the weak-coupling regime, they still exhibit substantial deviations in the strong-coupling regime $|g|\gtrsim 1.0$ MeV. In contrast, the FCIQMC results are in remarkable agreement with the exact energies across the entire range of $g$, with statistical errors being almost negligible. The benchmarking shows the superior accuracy of FCIQMC, especially in strongly coupled situations where truncated many-body methods break down (see Fig.~\ref{fig:fig_pairing}). Our implementation for the Richardson model is made publicly available for reproducibility and further development~\cite{githubcode}.

We now turn to much more challenging calculations of the realistic system of infinite nuclear matter. The many-body Hamiltonian, $\hat{H}=\hat{T}+\hat{V}_{\mathrm{NN}}+\hat{V}_{\mathrm{3N}}$, includes the kinetic energy term $\hat{T}$ alongside interaction terms corresponding to nucleon-nucleon (NN) and three-nucleon forces (3NF) that are derived from $\chi$EFT. To compare with other methods, the 3NF is included at the normal-ordered two-body level~\cite{PhysRevC.82.014314,PhysRevC.89.014319}, although FCIQMC can naturally deal with the full 3NF. The many-body basis is constructed by Slater determinants built on single-particle (SP) momentum eigenstates, $\left|\boldsymbol{k}_a \sigma_a \tau_a\right\rangle$ on the lattice in momentum space with $\boldsymbol{k}_a$,  $\sigma_a$ and $\tau_a$ being the momentum, spin and isospin projections, respectively. Our calculations are performed in a cubic box of volume $V=L^3$ with $A$ nucleons at a density $\rho=A/V$, employing periodic boundary conditions which discretize the momenta with a spacing of $\Delta k=2\pi/L$. The SP basis is defined by a momentum cutoff $|\boldsymbol{k}_a|\leq k_\mathrm{max}$, which should be sufficiently large to ensure convergence of solutions. Two distinct families of $\chi$EFT forces at the next-to-next-to-leading-order (N$^2$LO) are used: N$^2$LO(H{\"u}ther)~\cite{HUTHER2020135651} and $\mathrm{\Delta N^2LO_{GO}}$~\cite{PhysRevC.97.024332,PhysRevC.102.054301}, where the latter explicitly includes the $\Delta(1232)$ isobar degree of freedom.

\begin{figure*}[t]
    \centering
    \includegraphics[width=1\textwidth]{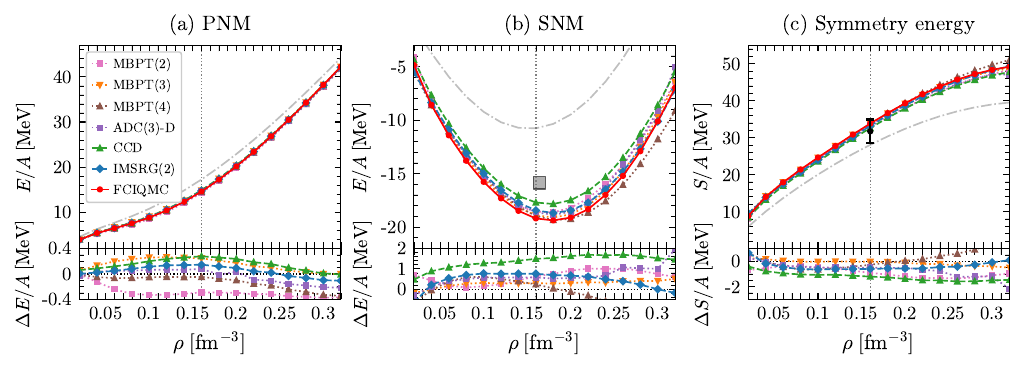}
    \caption{Energies per particle of PNM, SNM, and the nuclear symmetry energy $S(\rho)$ per nucleon as functions of density $\rho$. Calculations use the $\Delta$-full $\mathrm{\Delta N^2LO_{GO}}(450)$ interaction. The FCIQMC results are compared with MBPT(2), MBPT(3), MBPT(4), ADC(3)-D, CCD and IMSRG(2) in the same model space. The Hartree-Fock solutions are marked by dash-dot curves. The empirical saturation density is marked by vertical gray lines. Bottom panels show the energy differences of other methods relative to our FCIQMC results, highlighting the contribution of neglected correlations.}
    \label{fig:eos_go}
\end{figure*}

Before proceeding to large-scale calculations with FCIQMC, we perform a rigorous benchmark for SNM in a small model space where exact results can be obtained via direct diagonalization, compared with MBPT and IMSRG(2) methods. The results are shown in Fig.~\ref{fig:comparision}. Across all densities and for both families of nuclear forces, FCIQMC calculations with $10^5$ walkers reproduce exact energies with extraordinary accuracy, with deviations being orders of magnitude smaller than other methods. In contrast, both MBPT and IMSRG(2) exhibit significant discrepancies that grow substantially with increasing density and particularly for the harder interactions. MBPT(3) yields reasonable agreement in some cases, which depends on the interaction being used. This benchmark not only validates our FCIQMC implementation for infinite nuclear matter but also provides a clear illustration of the limitations of these widely-used truncated schemes for infinite nucleonic matter.

Next, we present large-scale calculations with the $\Delta$-full interaction, $\mathrm{\Delta N^2LO_{GO}}$ (450 MeV cutoff) \cite{PhysRevC.102.054301}, which has been successful in describing binding energies and radii of finite nuclei up to $^{132}$Sn~\cite{Koszorus2021,PhysRevLett.129.132501,PhysRevLett.130.032501,PhysRevLett.131.022502,PhysRevLett.132.162502,PhysRevLett.134.063002} but whose performance for nuclear matter saturation has been debated~\cite{marino2026,PhysRevC.109.059901,marino2024qdu}. We compare FCIQMC results with those from other six many-body methods, namely MBPT(2), MBPT(3), \textcolor{black}{MBPT(4)}, CCD, ADC(3)-D and IMSRG(2). All the calculations were performed in the same model space which contains 66 (76) nucleons and 682 (1364) SP states for PNM (SNM). The number of SP states is sufficiently large for the basis convergence. The uncertainty analysis can be found in the Supplemental Material~\cite{SM}.

As shown in Fig.~\ref{fig:eos_go}, for PNM, a relatively weakly correlated system, all methods yield similar energies per particle with differences below 0.5 MeV. For the SNM, however, the differences between FCIQMC and other methods become larger up to 2 MeV. Since SNM is more strongly correlated than PNM, the effect of the missing higher-order many-correlations is much more evident.
All methods overestimate both the saturation density and the saturation energy compared to the empirical saturation window $\rho_0=0.164\pm 0.007$ fm$^{-3}$ and $E_0/A=-15.86\pm 0.57$ MeV~\cite{PhysRevC.93.054314,PhysRevLett.122.042501}. Nevertheless, the resulting symmetry energies $S(\rho_0)$ are consistent with experimental constraints~\cite{Li2019}.

\begin{figure*}[t]
    \centering
    \includegraphics[width=1\textwidth]{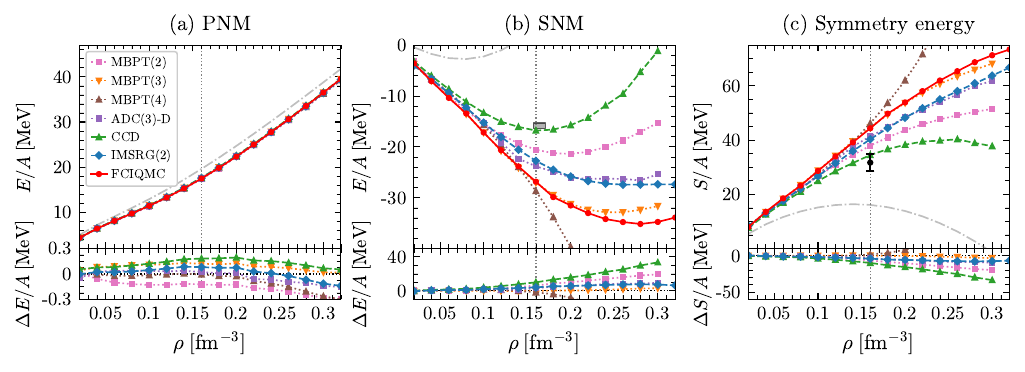}
    \caption{Similar to Fig.~\ref{fig:eos_go} but for calculations using the $\Delta$-less $\mathrm{N^2LO}$ interaction with cutoff 450 MeV from H\"uther \textit{et al}~\cite{HUTHER2020135651}. The discrepancies between FCIQMC and other truncated methods are much more evident for SNM.}
    \label{fig:eos_emn}
\end{figure*}

We also perform a parallel analysis with the $\Delta$-less N$^2$LO(H{\"u}ther) interaction~\cite{HUTHER2020135651} with the same cutoff of 450 MeV, which is able to reproduce experimental binding energies and charge radii of medium-mass nuclei up to $^{78}$Ni using IMSRG(2)~\cite{HUTHER2020135651}. Previous Brueckner-Hartree-Fock calculations concluded that this interaction fails to describe SNM saturation~\cite{PhysRevC.104.064312,MACHLEIDT2024104117,PhysRevC.104.034308}. However, this conclusion should also rely on many-body method being used.

Our results show significant differences in the binding energies of SNM among different methods with the $\Delta$-less N$^2$LO interaction. This is because this interaction is harder than the $\mathrm{\Delta N^2LO_{GO}}$ interaction. However, differences in PNM are still very small.
\textcolor{black}{Compared with Fig.~\ref{fig:eos_go}(b), the large differences seen in Fig.~\ref{fig:eos_emn}(b) indicate the stronger dependence on many-body method in SNM with this harder interaction. Though the MBPT(3) curve is close to FCIQMC, this is to some extent accidental, as demonstrated by MBPT(4) energies being significantly lower than MBPT(3) beyond the saturation density, indicating that the MBPT convergence is poor in high density SNM~\cite{drischler2026}. While the CCD curve appears to be close to the empirical saturation point, this can also be considered accidental, given that these results deviate from FCIQMC and that CCD is known to have limited accuracy in SNM~\cite{PhysRevC.89.014319}. The higher version CCD(T)~\cite{PhysRevC.110.054322} would likely be closer to FCIQMC. The IMSRG(2) shows a trend similar to ADC(3)-D. The relatively hard behavior of N$^2$LO(H{\"u}ther) in SNM may be related to its three-nucleon force, especially the rather large low-energy constant $c_\mathrm{D}=10.0$.}

Since FCIQMC incorporates correlations to all orders, the discrepancy between FCIQMC and other methods provides quantitative evidence for the crucial role of high-order many-body correlations. As illustrated in Fig.~\ref{fig:eos_emn}(b), the energy contribution from the missing correlations is about 10 MeV around saturation and about 40 MeV at 2.0$\rho_0$. This informs us that SNM is much more strongly-correlated than expected at higher densities. The results also demonstrated that the high-order correlations become strikingly important for hard nuclear forces. The high-order contribution is so large that it could be more significant than the $\chi$EFT truncation uncertainty~\cite{Epelbaum2015,PhysRevC.93.044002,PhysRevC.98.014002,q3vn-8y8s} at high densities. This finding underscores an important message to the field: to reliably assess the quality of a chiral interaction and to make credible progress in solving the long-standing saturation problem~\cite{PhysRevC.58.1804,LEJEUNE200045,PhysRevC.83.031301,MACHLEIDT2024104117,HEBELER20211}, it is necessary to improve the truncation schemes of existing post-Hartree-Fock methods. In this respect, FCIQMC is useful to disentangle the potential deficiencies of a nuclear Hamiltonian from the many-body approximations.

\textit{Summary.\textemdash}
In this Letter, we introduce the state-of-the-art FCIQMC method for \textit{ab initio} studies of infinite nuclear matter. By stochastically sampling in the complete many-body Hilbert space, FCIQMC can include full many-body correlations, circumventing the truncation errors of expansion methods like CC, IMSRG and ADC-SCGF. FCIQMC also provides a solution to the notorious sign problem in QMC methods and avoids the fixed-node biases of other QMC approaches. The wave functions are directly sampled in FCIQMC, facilitating the calculations of various observables.
This work establishes FCIQMC as a rigorous benchmarking tool for infinite nuclear matter, which is one of the most challenging strongly correlated systems in nature.

Our primary finding is that high-order many-body correlations, largely truncated in other methods, play a crucial role in describing nuclear matter saturation. For the hard chiral interaction such as  N$^2$LO(H{\"u}ther)~\cite{HUTHER2020135651}, the energy contribution from these correlations is found to be strikingly large up to 40 MeV at 2.0$\rho_0$, a magnitude comparable to the theoretical uncertainty of the chiral expansion itself, indicating that SNM is much more strongly correlated than expected. By eliminating many-body uncertainties, this work makes a substantial step in resolving the long-standing problem to simultaneously describe the bulk properties of finite nuclei and infinite nuclear matter from a single Hamiltonian. The FCIQMC as a robust tool to predict nuclear EoS also establishes a reliable link between microscopic nuclear forces and the astrophysical observables of neutron stars.

The capability of FCIQMC to controllably capture correlations to all orders opens new frontiers in \textit{ab initio} nuclear physics. It enables a clear separation of Hamiltonian versus many-body uncertainties, which is crucial for guiding the development of the next generation of high-precision interactions. Furthermore, applying FCIQMC to other frameworks will greatly advance \textit{ab initio} calculations of finite nuclei, nuclear reactions, and electroweak processes, with considerable impacts on our understanding of phenomena across nuclear physics and astrophysics.

\textit{Acknowledgments.\textemdash}
We are grateful to A. Alavi for valuable suggestions and discussions on the FCIQMC algorithm, and thank J. G. Li, S. Zhang, C. Barbieri and G. Col\`{o}  for useful discussions. This work has been supported by the National Key R\&D Program of China under Grant No. 2023YFA1606401 and 2024YFA1610900; the National Natural Science Foundation of China under Grants No. 12335007, 12535008 and 12475118, and the High-Performance Computing Platform of Peking University.
F.~Marino was supported by the Deutsche Forschungsgemeinschaft (DFG, German
Research Foundation) – Project-ID 279384907 – SFB 1245, and through the Cluster of Excellence “Precision Physics, Fundamental Interactions, and Structure of Matter” (PRISMA+ EXC 2118/1, Project ID 390831469). We acknowledge the CINECA awards AbINEF (HP10B3BG09) and RespGF (HP10BQMECT) under the ISCRA initiative, for the availability of highperformance computing resources and support. This work used the DiRAC Data Intensive service (DIaL3) at the University of Leicester, managed by the University of Leicester Research Computing Service on behalf of the STFC DiRAC HPC Facility (www.dirac.ac.uk). The DiRAC service at Leicester was funded by BEIS, UKRI and STFC capital funding and STFC operations grants. DiRAC is part of the UKRI Digital Research Infrastructure.

\textit{Data availability.\textemdash}The data that support the findings of this article and source code of FCIQMC for the pairing model are openly available~\cite{githubcode}.

\bibliographystyle{modified-apsrev4-2.bst}
\bibliography{reference}

\end{document}

\title{Supplemental Material for: \textit{Ab Initio} Exact Calculation of Strongly-Correlated Nucleonic Matter}

\author{R. Z. Hu\,\orcidlink{0009-0002-8797-6622}}
\affiliation{School of Physics, and State Key Laboratory of Nuclear Physics and Technology, Peking University, Beijing 100871, China}
\author{S. L. Jin\,\orcidlink{0000-0001-5958-8070}}
\affiliation{School of Physics, and State Key Laboratory of Nuclear Physics and Technology, Peking University, Beijing 100871, China}
\author{X. Zhen\,\orcidlink{0009-0000-1806-4123}}
\affiliation{School of Physics, and State Key Laboratory of Nuclear Physics and Technology, Peking University, Beijing 100871, China}
\author{H. Y. Shang\,\orcidlink{0009-0007-1253-4519}}
\affiliation{School of Physics, and State Key Laboratory of Nuclear Physics and Technology, Peking University, Beijing 100871, China}
\author{J. C. Pei\,\orcidlink{0000-0002-9286-1304}}\email[]{peij@pku.edu.cn}
\affiliation{School of Physics, and State Key Laboratory of Nuclear Physics and Technology, Peking University, Beijing 100871, China}
\affiliation{Southern Center for Nuclear-Science Theory (SCNT), Institute of Modern Physics, Chinese Academy of Sciences, Huizhou 516000, China}
\author{F. R. Xu\,\orcidlink{0000-0001-6699-0965}}\email[]{frxu@pku.edu.cn}
\affiliation{School of Physics, and State Key Laboratory of Nuclear Physics and Technology, Peking University, Beijing 100871, China}
\affiliation{Southern Center for Nuclear-Science Theory (SCNT), Institute of Modern Physics, Chinese Academy of Sciences, Huizhou 516000, China}
\author{F.~Marino\,\orcidlink{0000-0001-7743-1982}}
\affiliation{Institut f\"{u}r Kernphysik and PRISMA+ Cluster of Excellence, Johannes Gutenberg-Universit\"{a}t Mainz, 55128 Mainz, Germany}

\maketitle

\section{Algorithm Details}
In this section we provide all details of the FCIQMC algorithm in nuclear matter calculations~\cite{Booth18072014}.

\subsection{Master equation}
We consider the imaginary-time Schr\"odinger equation:
\begin{equation}
    -\dfrac{\mathrm{d}\ket{\Psi(\tau)}}{\mathrm{d}\tau} = (\hat{H}-E_0)\ket{\Psi(\tau)},
\end{equation}
where $\tau$ is the imaginary-time, $\hat{H}$ is the many-body Hamiltonian and $E_0$ is the ground-state energy.

Expanding the above equation in the full configuration space with $\ket{\Psi(\tau)}=\sum_i C_i(\tau)\ket{D_i}$, we obtain:
\begin{equation}
    -\dfrac{\mathrm{d}C_i(\tau)}{\mathrm{d}\tau} = \sum_j (H_{ij}-E_0\delta_{ij})C_j(\tau),
\end{equation}
where $H_{ij}=\braket{D_i}{\hat{H}|D_j}$ are complex-valued Hamiltonian matrix elements.

To solve the above equation stochastically, the so-called walkers are introduced in FCIQMC, which sample the wave function coefficients $C_i$ with the following definitions:
\begin{enumerate}
\item Each walker $a$ lives on a Slater determinant $\ket{D_i}$, which carries a ``direction'' $\hat{n}_a$ (a complex unit vector).
\item The total number of walkers of $\ket{D_i}$, denoted by $\hat{N}_{i}$, is defined by the summation of directions on $\ket{D_i}$:
\begin{equation}
    \hat{N}_{i}= \sum_{a\in\ket{D_i}} \hat{n}_a,
\end{equation}
\item The total walker number $N_{\mathrm{w}}$ is defined to be the summation of absolute value of $\hat{N}_{i}$:
\begin{equation}
    N_{\mathrm{w}}=\sum_i |\hat{N}_i|,
\end{equation}
\end{enumerate}

The resulting master equation for walkers in FCIQMC reads
\begin{equation}
    \label{equ:master_equation}
    \hat{N}_i(\tau+\Delta\tau) - \hat{N}_i(\tau) = -\Delta\tau(H_{ii}-S)\hat{N}_i(\tau) - \Delta\tau\sum_{j\neq i} H_{ij}\hat{N}_j(\tau),
\end{equation}
where $\Delta\tau$ is the time-step. So that wavefunctions are directly sampled in FCIQMC:
\begin{equation}
    \ket{\Psi(\tau)} = \sum_i \hat{N}_i(\tau) \ket{D_i}.
\end{equation}
In the $\tau\to\infty$ limit, the ground-state wave function $\ket{\Phi_0}$ is projected out:
\begin{equation}
    \ket{\Phi_0} \propto \lim_{\tau\to\infty} \ket{\Psi(\tau)}.
\end{equation}

\subsection{Evolution algorithm}

At each time-step $\Delta\tau$, we perform the following algorithm for all existing walkers:
\begin{enumerate}
\item \textit{diagonal step}: For each existing walker (on $\ket{D_i}$ with direction of $\hat{n}_i$), we first calculate its death probability:
    \begin{equation}
        p_\mathrm{death}(i) = \Delta\tau |H_{ii}-S|.
    \end{equation}
    If $H_{ii}-S>0$, this walker dies with a probability of $p_\mathrm{death}(i)$; otherwise ($H_{ii}-S<0$), this walker clones a new walker on $\ket{D_i}$ with the same direction with a probability of $p_\mathrm{death}(i)$.

\item \textit{spawning step}: For each existing walker (on $\ket{D_i}$ with direction of $\hat{n}_i$), we randomly choose a connected $\ket{D_f}$ with a probability of $p_\mathrm{gen}(f|i)$ through excitation generation algorithm (explained latter). This walker tries to spawn a new walker on $\ket{D_f}$ with the following probability:
    \begin{equation}
        p_\mathrm{s}(f|i) = \dfrac{\Delta \tau |H_{fi}|}{p_\mathrm{gen}(f|i)},
    \end{equation}
with the direction of the newly-spawned walker given by $-\hat{h}_{fi}\hat{n}_i$, where $\hat{h}_{fi}=H_{fi}/|H_{fi}|$.

\item \textit{annihilation step}: Collect all walkers (including both existing and newly-spawned walkers) on the same determinant $\ket{D_i}$ by summing all directions, resulting number of $\hat{N}_i$ walkers with the same direction $\hat{n}_i$:
    \begin{equation}
        \hat{N}_i=\sum_{a\in\ket{D_i} }\hat{n}_{a}=N_i \hat{n}_i.
    \end{equation}
\end{enumerate}

The above algorithm ensures that the stochastic evolution of walkers aligns with the master equation. The use of complex-valued walker numbers $\hat{N}_i$ is the key in treating complex Hamiltonians, which has been illustrated in Ref~\cite{Booth2013}.

\subsection{Excitation generation}
Another key algorithm of FCIQMC is the ``excitation generation" in the spawning step: how to randomly generate a connected $\ket{D_f}$ from $\ket{D_i}$? Given the initial determinant $\ket{D_i}$, the excitation generation algorithm enters two possible modes randomly: single or double excite, with normalized probabilities:
\begin{equation}
    p_{\mathrm{single}}+p_{\mathrm{double}}=1.
\end{equation}

\begin{enumerate}
\item \textit{single excite}: We randomly select an occupied orbit $\alpha_a$ from $\ket{D_i}$ with uniform probabilities $1/C_A^1$, where $A$ is the nucleon number. Then we find the one-body channel of $\alpha_a$ (defined by the conservation of momentum and isospin projection), and count the number of unoccupied orbits in this channel, giving $N(b|a)$. We randomly select another unoccupied orbit $\alpha_b$ from this set with uniform probabilities $1/N(b|a)$, and get the final determinant $\ket{D_f}$ by the excitation $\alpha_a\to\alpha_b$. In this case, the conditional probability of the excitation is
\begin{equation}
p_{\mathrm{gen}}^{-1}(f|i)=p_{\mathrm{single}}^{-1} C_A^1 N(b|a).
\end{equation}

\item \textit{double excite}: We randomly select two occupied orbits $(\alpha_a,\alpha_b)$ from $\ket{D_i}$ with uniform probabilities $1/C_A^2$. Then we find the two-body channel of $(\alpha_a,\alpha_b)$, and count the number of unoccupied two-body states in this channel, giving $N(cd|ab)$. We randomly select another unoccupied two-body state $(\alpha_c,\alpha_d)$ from this set with uniform probabilities $1/N(cd|ab)$, and get the final determinant $\ket{D_f}$ by the excitation $(\alpha_a,\alpha_b)\to(\alpha_c,\alpha_d)$. In this case, the conditional probability of the excitation is
\begin{equation}
p_{\mathrm{gen}}^{-1}(f|i)=p_{\mathrm{double}}^{-1} C_A^2 N(cd|ab).
\end{equation}
\end{enumerate}

We want to point out that the final results are independent with values of $p_\mathrm{single}$ and $p_\mathrm{double}$, which will be tested later.

\subsection{Evolution periods}

One practical FCIQMC calculation consists of the following periods:
\begin{enumerate}
\item \textit{Initialization}: Initialize the wave function by placing $N_{\mathrm{ini}}$ walkers on the lowest determinant $\ket{D_0}$ with the same direction such as $\hat{1}$. Typically we use $N_{\mathrm{ini}}=10$, which doesn't change the final results. One could also use a better initial wavefunction to accelerate the projection.
\item \textit{Warm-up period}: We perform the evolution algorithm as each time-step with the fixed-shift mode, where the shift is kept at a constant value $S=\braket{D_0}{\hat{H}|D_0}>E_0$, so that the total walker number increases exponentially until reaching a target number $N_\mathrm{w}$.
\item \textit{Projection period:} We perform the evolution algorithm as each time-step with the varying-shift mode, where the shift is self-updated every $A$ steps by:
\begin{equation}
S(\tau) = S(\tau - A\Delta \tau) - \dfrac{\xi}{A\Delta \tau} \ln{\dfrac{N_\mathrm{w}(\tau)}{N_\mathrm{w}(\tau-A\Delta \tau)}},
\end{equation}
which keeps the total walker number stable. We use an optimal choice of $A=10$ and $\xi=0.1$ from previous works~\cite{10.1063/1.3193710,10.1063/1.3681396}, and the final results show no dependence on these parameters. In this period, the wavefunction is projected onto the ground state progressively.
\item \textit{Statistical period}: After enough steps, $N_\mathrm{w}(\tau)$ and $S(\tau)$ will become stable, indicating the evolution reaches equilibrium and wavefunction has been projected onto the true ground state. Afterwards the statistical period begins. We continue the evolution for enough steps to perform statistical analysis for the ground-state energy. The shift $S$ can be used as an estimator for the ground-state energy, but we usually use the projected estimator because its statistical uncertainty is usually smaller:
\begin{equation}
E(\tau) = \dfrac{\braket{D_0}{\hat{H}|\Psi(\tau)}}{\braket{D_0}{\Psi(\tau)}} = \sum_{i} H_{0i} \dfrac{\hat{N}_i(\tau)}{\hat{N}_0(\tau)},
\end{equation}
where $\hat{N}_0(\tau)$ is the walker number on $\ket{D_0}$ and $H_{0i} = \braket{D_0}{H|D_i}$.
\end{enumerate}

\begin{figure*}[hbtp]
    \centering
    \includegraphics[width=0.82\textwidth]{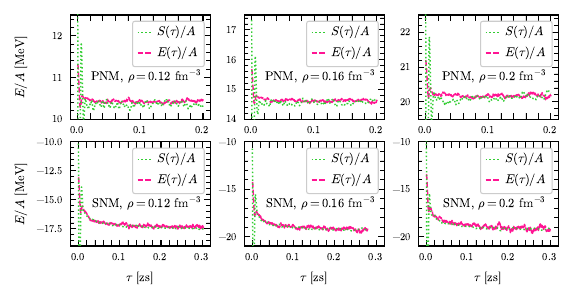}
    \caption{FCIQMC evolutions for PNM and SNM at three typical densities using $\Delta\mathrm{N^2LO_{GO}}(450)$ interaction. The calculations were performed in a model-space of $A=66\;(76)$ nucleons and $M=682\;(1364)$ single-particle states. We use $N_\mathrm{w}=2\times 10^7$ for PNM and $N_\mathrm{w}=2\times 10^8$ for SNM.}
    \label{fig:evolution}
\end{figure*}

In Fig.~\ref{fig:evolution}, we show the evolutions of $S(\tau)$ and $E(\tau)$ starting from the projection period in large-space calculations. In all calculations, the imaginary parts of $E/A$ are always very small (fluctuating around zero with small magnitudes less than 0.01 MeV), so we only plot the real parts of $E/A$.

\subsection{Initiator approximation}
We use the initiator approximation (i-FCIQMC) to suppress the sign problem during the evolution, which has been widely used in previous works~\cite{10.1063/1.3624383,PhysRevLett.109.230201,PhysRevB.85.081103,PhysRevLett.114.033001,PhysRevLett.117.115701}. In this adaption, a determinant $\ket{D_i}$ is defined to be an initiator determinant if $|\hat{N}_i|\geq n_\alpha$, where $n_\alpha$ is called the initiator threshold. During the evolution algorithm, we restrict walkers on non-initiator determinants from spawning to unoccupied determinants. This approximation introduces a dynamical truncation for the Hamiltonian:
\begin{equation}
\tilde{H}_{ij} =
\begin{cases}
0, & |\hat{N}_j|< n_\alpha \text{ and } \hat{N}_i=0,\\
H_{ij}, & \text{otherwise.}
\end{cases}
\end{equation}

During the evolution, the initiator determinants are self-adjusted dynamically at each step. This approximation naturally make use the sparse nature of the realistic nuclear many-body Hamiltonian, but brings a systematic bias to the calculation. However, in the large walker number limit $N_\mathrm{w}\to\infty$, all determinants will become initiators and this systematic bias is eliminated. In short, \textit{the only systematic uncertainty of FCIQMC method is originated from initiator approximation and can be well controlled by increasing total walker number}, which will be analyzed in detail later.

\subsection{Adaptive shift method}
The initiator bias can be efficiently corrected by using the adaptive shift method (AS-FCIQMC)~\cite{10.1063/1.5134006,10.1063/5.0005754}. In the adaptive shift method, each non-initiator determinant $\ket{D_i}$ has its own local shift $S_i(\tau)$ as a fraction of the full shift $S(\tau)$:
\begin{equation}
S_i(\tau) = f_i S(\tau),
\end{equation}
where the fraction $f_i$ is computed by monitoring the accepted and rejected spawning attempts due to the initiator criterion:
\begin{equation}
f_i = \dfrac{\sum_\mathrm{accepted}w_{ij}}{\sum_\mathrm{all}w_{ij}}.
\end{equation}
The weights $w_{ij}$ are derived from second-order perturbation theory:
\begin{equation}
w_{ij} = \dfrac{|H_{ij}|}{H_{jj}-E_0},
\end{equation}
where the unknown $E_0$ is replaced by $S$ in real calculations.

This approach can significantly accelerate the convergence of i-FCIQMC, which will also be tested later. Consequently, AS-FCIQMC is employed to calculate the equation of state (EoS) of PNM and SNM in this work.

\section{Uncertainty analysis}
The uncertainty in our FCIQMC calculation for the EoS of nuclear matter consist of five parts:
\begin{enumerate}
\item Chiral truncation: The theoretical uncertainty arising from the truncation of the chiral effective field theory ($\chi$EFT) interaction at a finite order.
\item Finite-size effects: Errors resulting from using a finite number of nucleons.
\item Model-space truncation: Errors due to the finite number of single-particle states.
\item Statistical uncertainty: The stochastic errors intrinsic to the Monte Carlo sampling process.
\item Initiator bias: The systematic error introduced by the initiator approximation, which depends on the total walker population.
\end{enumerate}
The first part is not discussed because the present work mainly focuses on many-body methods rather than the convergence of $\chi$EFT expansion. The other four parts are analyzed term by term in the following.

\subsection{Finite-size effects}
\begin{figure*}[hbtp]
    \centering
    \includegraphics[width=0.95\textwidth]{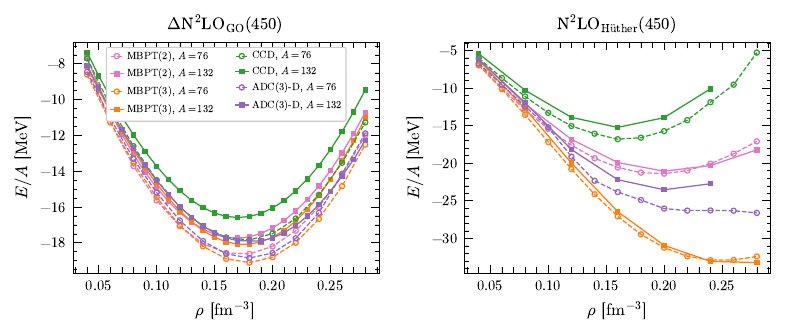}
    \caption{Finite-size effects of SNM EoS using 76 or 132 nucleons. The $\Delta\mathrm{N^2LO_{GO}}(450)$ and $\Delta$-less N$^2$LO$_\mathrm{H\ddot{u}ther}$(450) interaction are used.}
    \label{fig:finite_size_eos}
\end{figure*}

In our calculations, the period boundary condition is used, and the number of nucleons should be increased until the results converge to the thermodynamic limit. However, it is unrealistic to use a very large nucleon number in the \textit{ab initio} calculations because the computational cost increases very quickly with nucleon number. Luckily it is found that by using $A=66 \;(132)$ nucleons for PNM (SNM), the finite-size effects can be kept under control~\cite{PhysRevC.89.014319}. In particular, the kinetic energy remains close to that in the thermodynamic limit. This choice of nucleon numbers is state-of-the-art in existing \textit{ab initio} calculations for nuclear matter. In this work, we use $A=66$ for large-scale PNM calculations. However, the $A=132$ calculations for SNM are too computational demanding for FCIQMC at this time, requiring about $2\times 10^6$ CPU hours for the calculation for a single density. So in this work, we use $A=76$ nucleons for SNM.
As shown in Fig~\ref{fig:finite_size_eos}, by comparing results obtained with the less demanding MBPT(2), MBPT(3), CCD and ADC(3)-D methods, we estimate that \textcolor{black}{the deviations between $A=76$ and $A=132$ are about 0.6--1.6 MeV per particle for both interactions around the saturation density of $\rho=0.16$ fm$^{-3}$.}
\textcolor{black}{We expect a similar finite-size effect for FCIQMC. As commented in Refs.~\cite{PhysRevC.89.014319,PhysRevC.110.054322,marino2026}, the finite-size effect in nuclear matter with periodic boundary conditions mainly originates from the particle number, shell structure and momentum discretization, and is not very sensitive to the particular many-body method used. This expectation is supported by MBPT(2), MBPT(3), CCD and ADC(3)-D calculations given in Fig.~\ref{fig:finite_size_eos}, showing similar $A$ dependence.}
The twist-averaging technique has shown to be a practical way to remove finite-size effects~\cite{PhysRevE.64.016702}, which will be studied in our future work.

\subsection{Model space convergence}
\begin{figure*}[hbtp]
    \centering
    \includegraphics[width=0.7\textwidth]{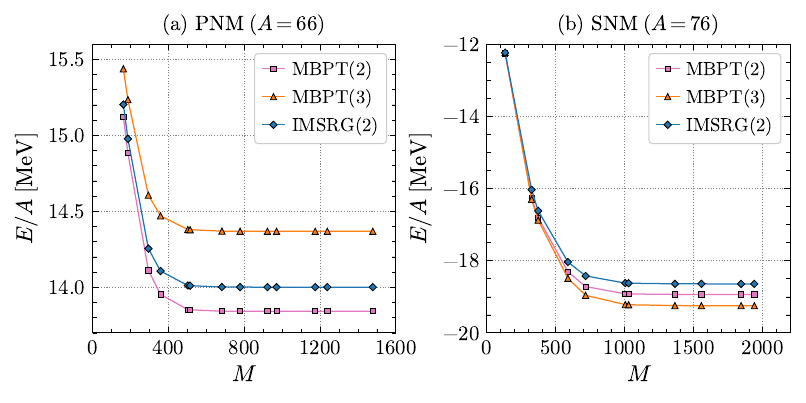}
    \caption{Model-space convergence of the ground-state energy per nucleon as a function of the total number of single-particle states $M$ for PNM and SNM. The calculations were performed using the $\Delta\mathrm{N^2LO_{GO}}(450)$ interaction at density $\rho=0.16\;\mathrm{fm^{-3}}$.}
    \label{fig:model_space}
\end{figure*}

Given the nucleon number $A=66\;(76)$, the model space convergence is examined by increasing the number of single-particle states, $M$, which is defined by $n_x^2+n_y^2+n_z^2\leq N_\mathrm{max}^2$. As shown in Fig.~\ref{fig:model_space}, a basis size of $N_\mathrm{max}^2=18$, $M=682\;(1364)$ is sufficient to ensure the model space convergence for PNM (SNM) calculations, which is used to calculate the EoS in this work. It is worth noting that the convergence behavior with respect to the model space size is largely independent of specific many-body method used. As a result, we use MBPT(2), MBPT(3) and IMSRG(2) to determine the necessary model space size, because FCIQMC calculations are much more computationally demanding. We have also checked that the model space convergence for the $\Delta$-less $\mathrm{N^2LO}(450)$ interaction used in this work is also similar.

\subsection{Statistical analysis}
In FCIQMC calculations, the sampled observables exhibit intrinsic autocorrelations. Consequently, a direct calculation of the standard error from raw data would lead to underestimation of the true statistical uncertainty. To obtain a reliable error estimate, the reblocking analysis technique~\cite{10.1063/1.457480,pyblock} is used. Taking the observable $X$ as an example, the procedure is described as follows:

\begin{enumerate}
\item \textit{Data partitioning}: A set of $N$ correlated samples $\{X_i\}$ is partitioned into $N_L$ contiguous blocks, each containing $L$ sequential data points.
\item \textit{Block averaging}: For each block $i$, the local average $\bar{X}_i(L)$ is computed.
\item \textit{Statistical estimation}: The ensemble mean $\bar{X}(L)$ and the standard deviation of the block means $\sigma(L)$ are determined from the $N_L$ blocks. The estimated statistical error $\Delta\bar{X}(L)$ and its associated uncertainty $\delta\Delta\bar{X}(L)$ are given by:
\begin{equation}
\Delta\bar{X}(L) = \frac{\sigma(L)}{\sqrt{N_L - 1}},
\end{equation}
\begin{equation}
\delta\Delta\bar{X}(L) = \frac{\Delta\bar{X}(L)}{\sqrt{2(N_L - 1)}}.
\end{equation}
\item \textit{Convergence}: By systematically increasing the block size (typically as $L = 2^n$), we monitor the evolution of $\Delta\bar{X}(L)$. The statistical error is identified as the plateau value where $\Delta\bar{X}(L)$ converges, indicating that the blocks have become effectively uncorrelated.
\end{enumerate}

\begin{figure*}[hbtp]
    \centering
    \includegraphics[width=0.47\textwidth]{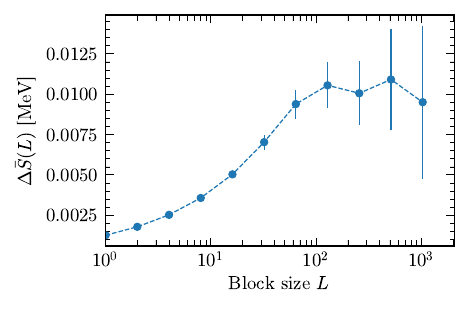}
    \caption{Blocking analysis for the shift $S$ in a FCIQMC calculation. The circles denote the estimated statistical error, $\Delta\bar{S}(L)$, and the error bars represent the uncertainty in the estimate, $\delta\Delta\bar{S}(L)$.}
    \label{fig:analysis}
\end{figure*}

Figure~\ref{fig:analysis} shows a representative blocking analysis for $S$ in a realistic FCIQMC calculation. The results indicate that a block size of $L\simeq 2^{10}$ is sufficient to give a converged estimation for statistical error.

\subsection{Initiator bias}
The dominant source of systematic uncertainty in our FCIQMC calculations is the initiator bias, which can be systematically eliminated by increasing $N_\mathrm{w}$, towards the FCI limit. The $N_\mathrm{w}$ convergence can be established based on two criteria: (1) The calculated energy reaches a plateau as $N_\mathrm{w}$ increases, and (2) The converged results exhibit independence from the initiator threshold $n_\alpha$.

\begin{figure*}[hbtp]
    \centering
    \includegraphics[width=0.72\textwidth]{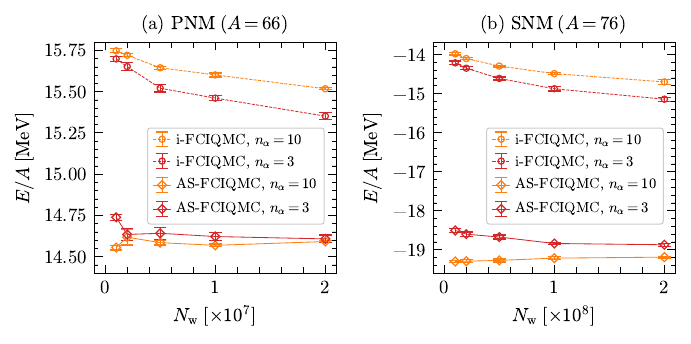}
    \caption{Convergence of the ground-state energy per nucleon as a function of walker number $N_\mathrm{w}$ for different initiator thresholds $n_\alpha$, using standard initiator and adaptive shift methods. All calculations were performed using the $\Delta\mathrm{N^2LO_{GO}}(450)$ interaction at density $\rho=0.16\;\mathrm{fm^{-3}}$.}
    \label{fig:convergence_nw}
\end{figure*}

We have performed calculations using both the standard i-FCIQMC and the improved AS-FCIQMC methods. As shown in Fig.~\ref{fig:convergence_nw}, the AS-FCIQMC method demonstrates significantly improved convergence properties. Specifically, for PNM, the AS-FCIQMC results are fully converged with $N_\mathrm{w}=2\times 10^7$ walkers, while the standard i-FCIQMC calculations fail to reach convergence within the same range of walker populations.

For the more challenging SNM case, we observe a bracketing behavior: the AS-FCIQMC results with $n_\alpha=3$ converge from above, while those with $n_\alpha=10$ converge from below. Consequently, we can constrain the exact result enveloped between these two curves, with an estimated residual initiator bias of around 1.5\%. Since the calculations with $n_\alpha=10$ usually carry smaller statistical uncertainties than those with $n_\alpha=3$, we adopt $n_\alpha=10$ to calculate the EoS for PNM and SNM in this work, providing a lower-bound for the exact result. It is worth noting that this systematic bias can be further reduced by using larger $N_\mathrm{w}$. These results also highlight the necessity of the adaptive-shift technique for such strongly correlated systems. Similar techniques have been widely used in selected CI methods in quantum chemistry~\cite{10.1063/1.4869192,10.1063/1.1679199,10.1063/1.4992127,Sharma2017,Holmes2016}.

\subsection{Independence on other parameters}
\begin{figure*}[t]
    \centering
    \includegraphics[width=0.45\textwidth]{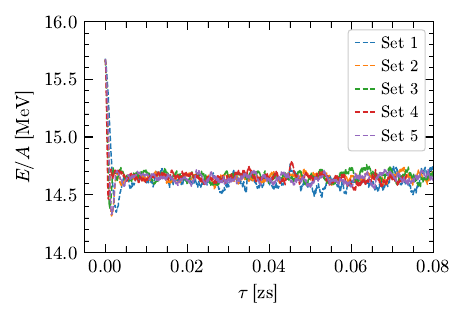}
    \caption{Imaginary-time evolutions for the different parameter sets listed in Table~\ref{table:params_test}.}
    \label{fig:test}
\end{figure*}

The FCIQMC results are independent of specific choice of other parameters, such as the imaginary-time step $\Delta\tau$, and the probabilities $p_\mathrm{single}$ and $p_\mathrm{double}$ in excitation generation algorithm. To verify this, we have performed parallel calculations for PNM using $\Delta\mathrm{N^2LO_{GO}}(450)$ interaction at 0.16 fm$^{-3}$ density, with five parameter sets shown in Table~\ref{table:params_test}. Sets 1 through 4 test the dependence on the time step, while Set 5 examines the effect of varying the excitation generation probabilities. As shown in Fig.~\ref{fig:test}, the imaginary-time evolution of the energy remains consistent across all sets. It is evident that variations in these parameters may slightly affect the sampling efficiency (reflected in the statistical uncertainties) but do not systematically shift the estimated ground-state energy.

\begin{table}[t]
    \centering
    \caption{Sensitivity check of the ground-state energy per nucleon with respect to the time step and excitation probabilities. $\bar{E}/A$ is the calculated mean value and $\Delta \bar{E}/A$ is the corresponding statistical uncertainty.}
    \renewcommand{\arraystretch}{1.45}
    \setlength{\tabcolsep}{2.0mm}
    \begin{tabular}{cccccc}
    \hline\hline
    Set & $\Delta\tau$(zs) & $p_\mathrm{single}$ & $p_\mathrm{double}$  & $\bar{E}/A$ (MeV) & $\Delta \bar{E}/A$ (MeV)\\
    \hline
    1 & $5\times 10^{-6}$ & 0.1 & 0.9 & 14.6252 & 0.030\\
    2 & $2\times 10^{-6}$ & 0.1 & 0.9 & 14.6581 & 0.012\\
    3 & $1\times 10^{-6}$ & 0.1 & 0.9 & 14.6610 & 0.014\\
    4 & $5\times 10^{-7}$ & 0.1 & 0.9 & 14.6397 & 0.010\\
    5 & $2\times 10^{-6}$ & 0.3 & 0.7 & 14.6401 & 0.012\\
    \hline\hline
\end{tabular}
\label{table:params_test}
\end{table}

\subsection{Computational cost and future improvements}
\textcolor{black}{For a given particle number, density, and single-particle model space, the computational cost of FCIQMC is approximately proportional to the total walker population $N_\mathrm{w}$, since the diagonal and spawning steps are performed walker by walker and the annihilation step scales with the number of spawned and existing walkers. The required $N_\mathrm{w}$ is determined by estimating residual initiator bias. For the $A=132$ SNM calculations with the present nuclear interactions and model spaces, we estimate that a walker population of the order of $N_\mathrm{w}\sim 10^9$ would be required to reduce the FCIQMC uncertainty to a well-controlled level.}

\textcolor{black}{Our present FCIQMC code for nuclear matter is written in C++17 and uses OpenMP and MPI for shared-memory parallelism and distributed-memory communication. For an $A=132$ SNM calculation with $N_\mathrm{w}\sim 10^9$, the current implementation would require roughly $10^6$--$10^7$ CPU hours for each density point. Although increasing the particle number is a direct way to reduce finite-size effects under periodic boundary conditions, it is not the most efficient route within FCIQMC framework. A more practical strategy is to use twist-averaged boundary conditions~\cite{PhysRevE.64.016702}, which can reduce shell and finite-size effects by averaging over boundary phases without requiring a comparably large increase in $A$.}

\textcolor{black}{There is, however, substantial room for improving the efficiency of FCIQMC calculations. The excitation generator used in the present work samples allowed single and double excitations uniformly within symmetry-conserving channels. More efficient generators can instead sample connected determinants with probabilities that reflect the magnitudes of the Hamiltonian matrix elements. Related heat-bath and selected-CI ideas have been proposed in quantum chemistry to greatly improve the quality of excitation generation~\cite{Holmes2016,Sharma2017}, and precomputed or partially precomputed heat-bath generators can be adapted to nuclear-matter Hamiltonians. Another promising direction is the semi-stochastic projector-QMC strategy, in which the imaginary-time projection is performed deterministically in a compact important subspace and stochastically in the remaining Hilbert space~\cite{PhysRevLett.109.230201,10.1063/5.0005754}. Such hybrid deterministic-stochastic algorithms can reduce statistical noise and initiator bias substantially, with reported efficiency gains of two to three orders of magnitude in many cases.}

\textcolor{black}{Finally, the structure of FCIQMC algorithm is well suited to modern computing accelerator architectures. Most walker spawning and death/cloning operations are local, while communication is primarily required by a single annihilation step. This locality makes a GPU implementation a natural route toward much larger walker populations. We therefore expect that a combination of improved excitation generation, semi-stochastic projection, twist-averaged boundary conditions, and GPU-oriented implementations will significantly reduce the remaining uncertainties of FCIQMC calculations for nuclear systems.}

\bibliographystyle{modified-apsrev4-2.bst}
\bibliography{reference}